\documentclass[aps,pre,onecolumn,superscriptaddress]{revtex4}

\newcommand{\eg}{{\it e.g.},~}
\newcommand{\ie}{{\it i.e.},~}

\usepackage{epsfig,xcolor,times}
\usepackage{siunitx}
\usepackage{amsmath, mathrsfs}
\graphicspath{{FIG/}{FIG/SUPPLEMENT/}}
\begin{document}

\title{The Shape of Hanging Elastic Cylinders}
\author{Serge Mora}
\email[]{serge.mora@umontpellier.fr}
\affiliation{Laboratoire de M\'ecanique et de G\'enie Civil, Universit\'e de Montpellier and CNRS.\\ 163 rue Auguste Broussonnet. F-34090 Montpellier, France.}

\author{Edward And\`o}
\affiliation{Laboratoire 3SR, Universit\'e Grenoble Alpes and CNRS. F-38041 Grenoble, France.}

\author{Jean-Marc Fromental}
\affiliation{Laboratoire Charles Coulomb, Universit\'e de Montpellier and CNRS.\\ 163 rue Auguste Broussonnet. F-34090 Montpellier, France.}

\author{Ty Phou}
\affiliation{Laboratoire Charles Coulomb, Universit\'e de Montpellier and CNRS.\\ 163 rue Auguste Broussonnet. F-34090 Montpellier, France.}

\author{Yves Pomeau}
\affiliation{University of Arizona, Department of Mathematics, Tucson, USA.}

\date{\today}
\begin{abstract}
Deformations of heavy elastic cylinders with their axis in the direction of earth's gravity field are investigated.
The specimens, made of polyacrylamide hydrogels, are attached from their top circular cross section to a rigid plate.
An equilibrium configuration results from the interplay between gravity that tends to deform the cylinders downwards under their own weight, and elasticity that resists these distortions.
The corresponding steady state exhibits fascinating shapes which are measured with lab-based micro-tomography.
For any given initial radius to height ratio, the deformed cylinders are no longer axially symmetric beyond a critical value of a control parameter that depends on the volume force, the height and the elastic modulus: self-similar wrinkling hierarchies develop, and dimples appear at the bottom surface of the shallowest samples. We show that these patterns are the consequences of elastic instabilities.
\end{abstract}

\maketitle

\section{Introduction}

Many materials such as biological tissues can withstand huge {\it elastic} deformations over more than several hundred percent. By {\it elastic} we mean that once the applied forces are released, the system spontaneously recovers its reference shape. Fascinating and puzzling shapes can result from these large deformations\cite{Hayward2010,Dervaux2012,Saintyves2013,Tallinen2014,Mora_prl2014,Chakrabarti2016} which have to be better understood in view of emerging applications as computer-assisted precision surgery involving human organs among which some are elastic and can undergo large deformations\cite{Cotin1999,Liu2003,Taylor2008}. A physical rationalization of these phenomena requires first to model materials subjected to basic external loads in simple geometries.

In this paper, we investigate deformations of elastic cylinders attached to a rigid plate at the upper cross section, and submitted to a downwards constant volume force in the vertical direction. This minimal system is enough to exhibit intriguing shapes such as those encountered in more complex systems with practical applications. 
We demonstrate that when the constant volume force (\eg  gravity or another acceleration) is switched on and a new equilibrium steady state is reached, the displacement of matter inside the cylinder is either divergent (the mass has moved toward the circumference of the cylinder) or convergent (towards the axis) depending on the initial slenderness of the cylinder. In addition, cascading and asymmetric wrinkles develop at the vertical surface beyond critical volume forces, and dimples appear along the lower surface of the shallowest cylinders.
We demonstrate that the mechanisms responsible for this variety of equilibrium shapes arise from elastic instabilities that can occur simultaneously.

\section{Materials and methods}

We use aqueous polyacrylamide gels prepared by copolymerization of acrylamide (concentration ranging from $\SI{28}{\gram \per \liter}$ to $\SI{46}{\gram \per \liter}$ depending on the sample) and
N,N'−methylenebisacrylamide (from $\SI{0.18}{\gram \per \liter}$ to $\SI{0.37}{\gram \per \liter}$)  in the presence of initiators, tetramethylenediamine ($\SI{0.6}{\gram \per \liter}$) and sodium persulfate ($\SI{0.93}{\gram \per \liter}$), in deionized water. Prior to mixing the constituents, all solutions are saturated with nitrogen gas, to ensure the near insufficiency of oxygen \cite{Menter2000}. Polymerization yields a loose permanent polymer network swollen in water (called hydrogel). The characteristic time scale of the diffusion of the solvent through the network is $\tau_d \simeq \ell^2/D_c$, with $D_c$ the cooperative diffusion coefficient and $\ell$ a characteristic length. For polyacrylamide gels $D_c\sim \SI{1e-11}{\square \meter \per \second}$ \cite{Hecht1981}. In what follows, we mainly deal with length scales larger than $\SI{1}{\milli\meter}$, hence $\tau_d > \SI{1e5}{\second}$, a time scale far larger than those of our experiment. Hence, the gels can be considered here as incompressible \cite{Hui2005,Lin2007}, with a mass density $\rho$ almost equal to that of water. In addition, they behave as an isotropic elastic material for strains up to several hundreds percent. The shear modulus $\mu$ can be tuned by varying the concentrations in monomers and cross-linkers.\\

In our experiments, the reagents generating the gels are first dissolved into ultra-pure water and poured into cylindrical dishes up to the brim.
The radius of the dishes, $R$, and the height, $h$, are between \SI{36}{\milli\meter} and \SI{150}{\milli\meter} and between \SI{8}{\milli\meter} and \SI{40}{\milli\meter} respectively.
The shear modulus of the gel is measured by indentation of a non-adhesive rigid sphere (diameter $\SI{6}{\milli\meter}$ at the surface of control samples fully covered with pure water in order to remove capillary forces \cite{Czerner2015,Tong2015,Arora2018}.
For the experiments reported here, it lies between 10 and \SI{160}{\pascal}, a low value compared to modulus of ordinary rubber-like materials ($\mu \sim$ \SI{1}{\mega\pascal}), but comparable with the modulus of tissues (as liver or brain).
After the gel is cured into the dish, its top surface is attached by capillary action onto a rigid horizontal surface coated with a sheet of paper, and the dish is gently removed from below. This results in a hanging cylinder of gel attached from the top and deformed by the gravitational acceleration $g$.
Samples with shear modulus as low as few tens of Pascals are highly sensitive to any perturbations during the sample handling.
Care is therefore required to avoid damage that would lead to changes in the equilibrium deformed shapes of the samples.   
The hanging cylinders spontaneously recover the size and shape of the moulds when immersed into pure water, whose mass density is equal to that of the gels. Moreover, they recover the same deformed shape after being withdrawn from water to air. These observations indicate that the deformations are reversible.
The dimensionless gravity acceleration $\alpha$ is defined as the ratio of the characteristic gravitational stress, $\rho g h$, to the shear modulus: $\alpha = \rho g h/\mu$. $\alpha$ lies between 1 and 16 in the experiments reported here.
Such large values of $\alpha$ could also be obtained with samples of higher shear modulus accompanied by larger heights or stronger accelerations.
Surface tension is not always negligible with soft solids \cite{Mora_prl2010,Mora_prl2013,Style2013,Jagota2014,Mora_JPhys2015}, and the dimensionless parameter $\alpha$ has to be complemented with another one, namely $\beta =\gamma/(\mu h)$, with $\gamma$ the surface energy per unit area. For polyacrylamide gels in air, $\gamma\simeq$ \SI{70}{\milli\newton\per\meter}. Taking for instance $\mu\sim$ \SI{35}{\pascal} and $h\sim$ \SI{20}{\milli\meter} as characteristic shear modulus and height, one gets $\beta\sim 0.1$. \\

The difference in x-ray absorption between air and water makes x-ray tomography a valuable tool to reveal the structure of these hydrogels in 3D.
The equilibrium shapes of the hanging cylinders are acquired using an RX-Solutions x-ray scanner with a Hamamastu L12161-07 micro-focus source and a Varian Paxscan DX2520 amorphous silica detector.
The machine is set to perform a continuous \SI{360}{\degree} rotation during tomography acquisition to ensure the sample does not move while scanning. To limit drying scanning time is reduced by accepting a degraded spatial resolution (using the detector in 2x2 binning) which allows scans in the 3-15 minutes time range.
Reconstructions yield a three dimensional field quantifying the x-ray attenuation of the object discretized into cubic voxels with size ranging from $(\SI{50}{\micro\meter})^3$ to $(\SI{200}{\micro\meter})^3$ depending on the size of the specimen. The grey levels of each axial slices have been averaged pixel by pixel over the azimuth (from 0 to 360 degrees by steps of 1 degree), yielding the mean vertical slice of the hanging cylinders as presented in Fig.~\ref{fig : profils} for six samples (vertical slices of other samples are shown in Sec.~\ref{sec : 2d}).

\section{Global deformation}

\begin{figure}[!h]
\begin{center}
\includegraphics[width=0.5\textwidth]{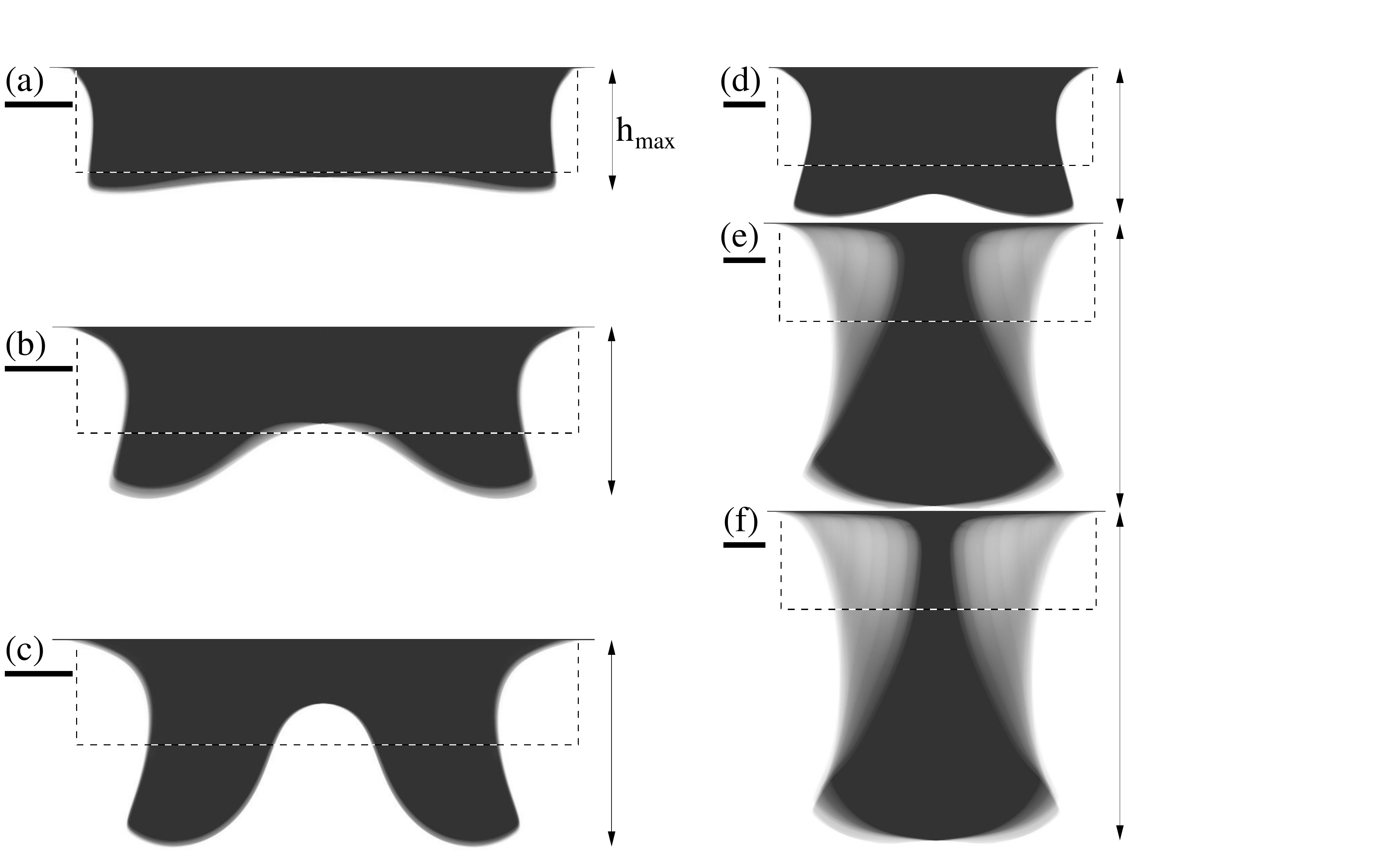}
\end{center}
\caption{Mean vertical slices of hanging elastic cylinders whose of initial domains are indicated by the dashed straight lines (bar length is $\SI{1}{\centi\meter}$), with dimensions $R\times h=\SI{37x15}{\milli\meter}$ for (a),(b),(c) and  $\SI{37x21}{\milli\meter}$ for (d),(e),(f). The gray level distributions in the mean vertical slices of (e) or (f) result from non-axially symmetric shapes.
 The shear modulus of the gels are $\mu= \SI{77.9\pm1.5}{\pascal}$ (a,d),$\SI{37.9\pm1.5}{\pascal}$ (b,e), and \SI{29.2\pm 1.5}{\pascal} (c,f). Vertical double arrows define $h_{max}$ for each sample.
 The dimensionless accelerations ($\alpha$) and surface tensions ($\beta$) are respectively ($1.89\pm 0.04,~0.06\pm 0.001$) for (a); ($3.88\pm 0.1,~0.12\pm 0.005$) for (b); ($5.03\pm 0.2,~0.16 \pm 0.01$) for (c);  ($3.02\pm 0.05,~ 0.037\pm 0.001$) for (d);  ($6.21\pm 0.2,~ 0.077\pm 0.003$) for (e); and  ($8.05\pm 0.4,~0.1 \pm 0.005$) for (f).
}
\label{fig : profils}
\end{figure}

The shapes of the two stiffer samples ((a) and (d)) shown in Fig.~\ref{fig : profils} share a common feature: the vertical stretching is higher at the periphery than near the axis of symmetry, yielding a shallow central depression which can be explained as follows:
the downwards movement of a material point located on the axis of the cylinder involves deformations over a larger volume of the elastic material than the downwards movement of a material point located near the outer vertical surface of the cylinder, hence the larger displacements of material points located near the outer vertical surface.
This is true for any aspect ratio in the limit of the small strains.
Let us consider now softer samples.
A central depression subsists and it is more pronounced in samples (b),(c). Indeed, in the large aspect ratio limit ($R\gg h$), the volume influenced by the lateral outer surface can be sketched as a hollow cylinder of thickness $h$, hence $h$ is the characteristic width of the hanging corona.
The depressions previously reported are replaced by a bump at the lower surface of samples (e) and (f). This is because the (initial) radius is not large enough compared to the (initial) height of these samples so that displacements of two material points diametrically opposed are not independent anymore but cooperative. These arguments are supported by numerical simulations (see Fig.~\ref{fig : numerics 2d}). 

\begin{figure}[!h]
\begin{center}
\includegraphics[width=0.5\textwidth]{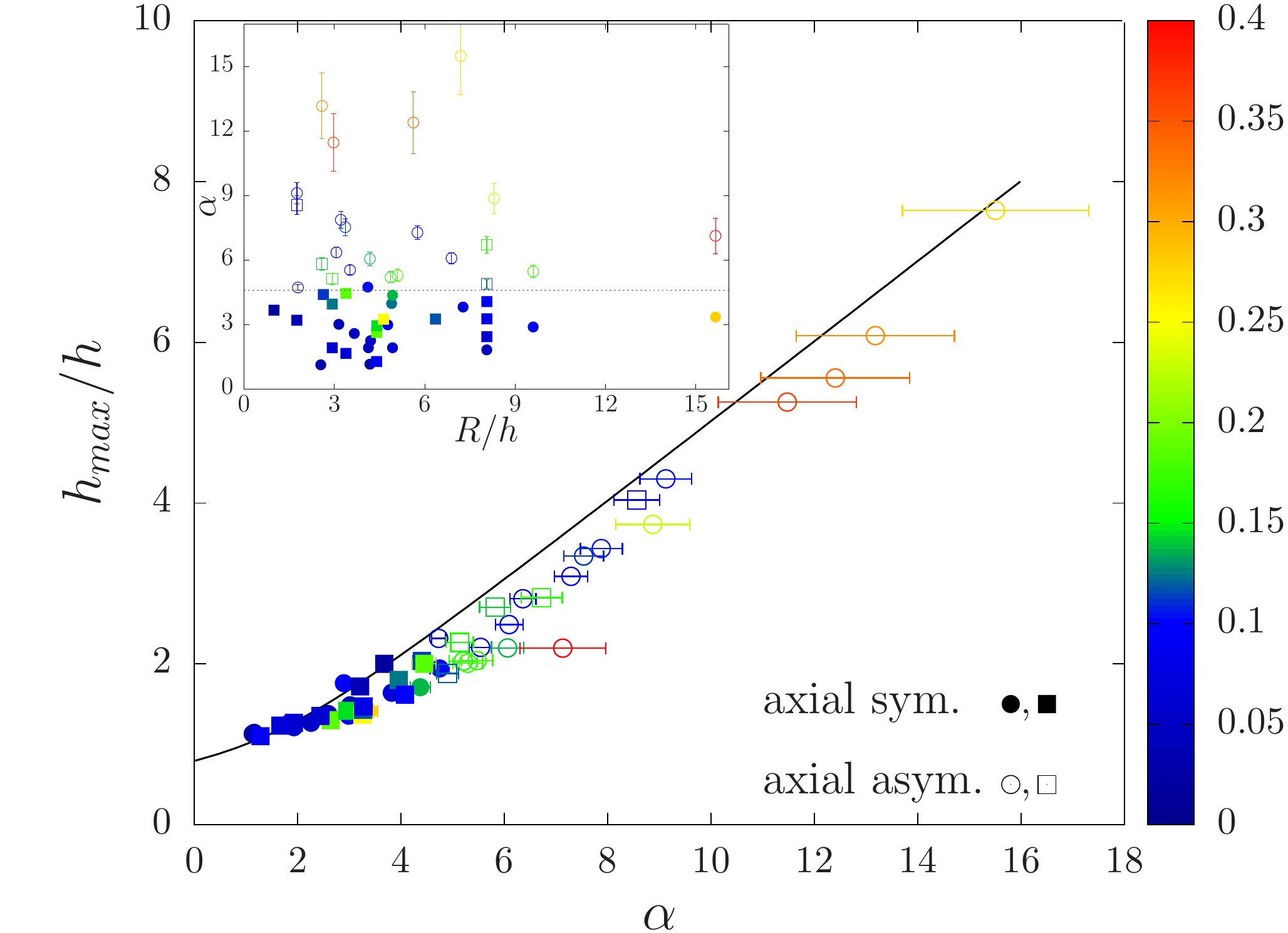}
\end{center}
\caption{Global stretch ratio $h_{max}/h$ as a function of the dimensionless acceleration $\alpha$. Filled symbols are for deformed samples with axial symmetry. Empty symbols are for deformed samples with a circumferential asymmetry. Colors indicate the value of the dimensionless surface tension $\beta$. Solid line is the prediction of Eq.~\ref{eqn : h_max}. Circles are for the samples acquired by micro-tomography, and squares for the other samples.
{\bf Inset:} $\alpha$ as a function of aspect ratio $R/h$ for all the tested samples. The dashed line, $\alpha=4.7$, shows that the demarcation line between axially symmetric and circumferential asymmetric deformed samples does not significantly depend on  $R/h$.} \label{fig : length}
 \end{figure}

In Fig.~\ref{fig : length} the global stretch ratio defined as the ratio of the equilibrium height of the hanging cylinder, $h_{max}$, to the initial height, $h$, is plotted as a function of $\alpha$.
These measurements are supplemented with a series of further samples that have not been acquired by micro-tomography (as indicated in  Fig.~\ref{fig : length}). The inset of Fig.~\ref{fig : length} shows, for the same set of samples, $\alpha$ as a function of $R/h$, underlining the range of aspect ratios in our experiments.
 The dimensionless surface energy $\beta$ being small compared to $\alpha$, the surface energy can be neglected in a first estimation of the total energy of the hanging cylinders, and we consider only the sum of gravity and elastic energies.
This later is estimated by using the strain energy density function of a isotropic and incompressible neo-Hookean solid, a model known to well describe polyacrylamide hydrogels.
This model is the natural extension of Hooke's law, valid for finite deformations. Depending on the aspect ratio of the reference shape of the cylinder, the deformation is either located at the border (shallow cylinders) or in the whole sample. Denoting ${\cal V}$ the volume involved in the deformation, the  Helmholtz free energy ${\cal E}$ can be expressed as: 
\begin{equation}
  \frac{{\cal E}}{{\cal V}}\sim \frac{{\mu}}{2}\left({\lambda}^2-\frac{2}{{\lambda}}\right)-{\rho g} \frac{h_{max}}{2}
  \label{eqn : Helmholtz}
\end{equation}
with  ${\lambda} \sim {\frac{h_{max}}{h}}$ a characteristic value of the strain. The first term in the right hand side in Eq.~\ref{eqn : Helmholtz} is the strain energy density, and the second term is the gravity energy density. The  Helmholtz free energy being minimal at equilibrium (${\partial {\cal E}}/{\partial h_{max}}=0$),
\begin{equation}
  2\left(h_{max}/h\right)^3 -\alpha \left(h_{max}/h\right)^2-1=0,
\label{eqn : h_max}
\end{equation}
$h_{max}/h$ computed from this equation is plotted in Fig.~\ref{fig : length}, with a good agreement with measurements. The measured $h_{max}/h$ is slightly smaller than the theoretical estimation due to the surface tension and the assumption that $h$ and ${\cal V}$ are not related. 

\begin{figure}[!h]
\begin{center}
\includegraphics[width=0.5\textwidth]{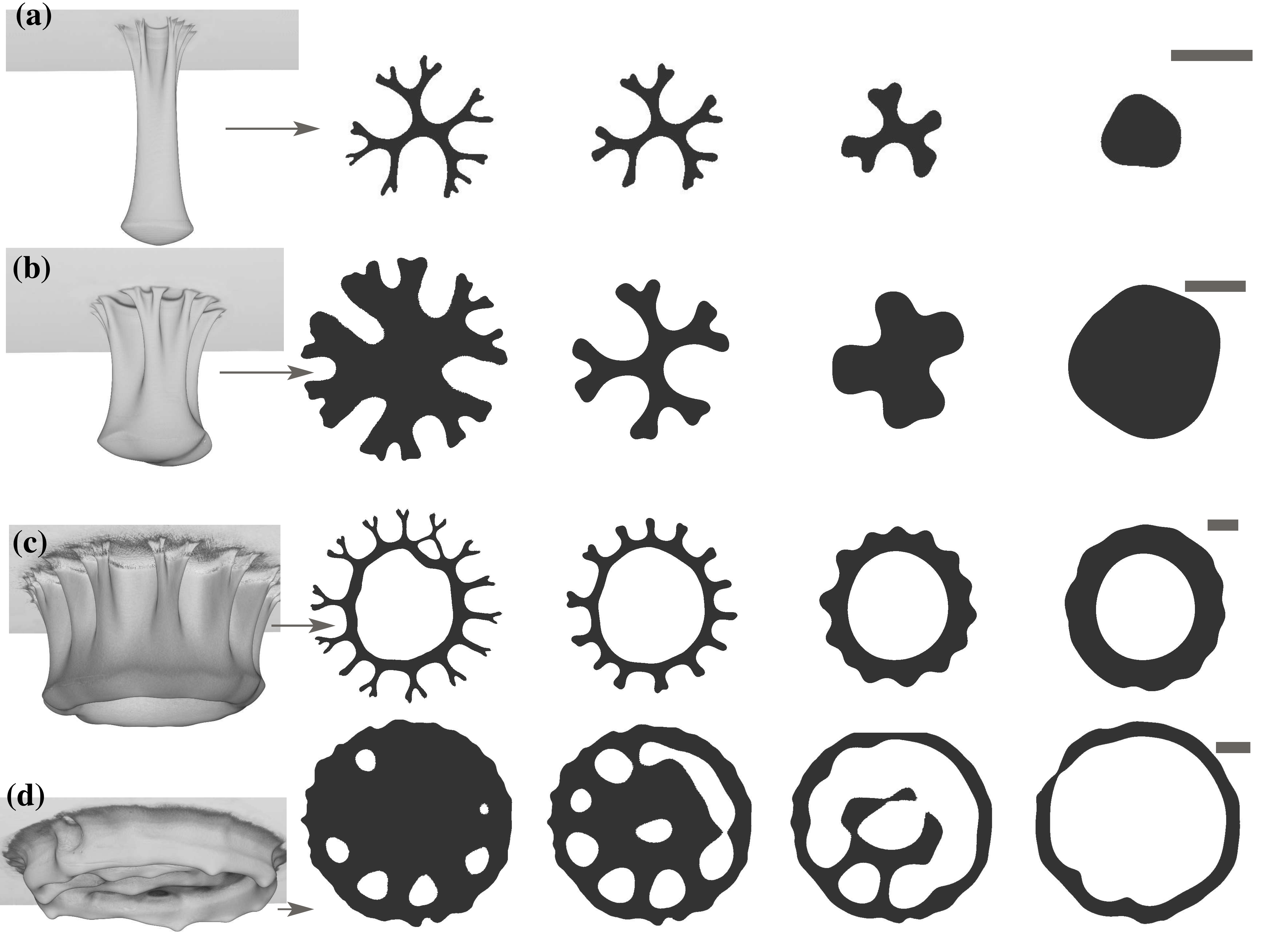}
\end{center}
\caption{Three-dimensional views and horizontal slices at various heights (from the top to the bottom of the samples) of 4 hanging cylinders with initial dimensions ($R\times h$) and shear modulus: \SI{22x25}{\milli\meter}, $\mu= $ \SI{27.4\pm1.5}{\pascal}  (a); \SI{37x24}{\milli\meter}, $\mu= $ \SI{29.2\pm 1.5}{\pascal}  (b); \SI{69x17}{\milli\meter}, $\mu= $ \SI{18.6\pm 1.5}{\pascal}  (c); \SI{72x9}{\milli\meter}, $\mu= $ \SI{12.9 \pm 1.5}{\pascal} (d).
    In the slices, the gel is in black; the white areas continuously surrounded by black areas correspond to the central depression (c,d) or to dimples (d).   
    The dimensionless accelerations ($\alpha$) and surface tensions ($\beta$) for these samples are respectively ($8.94\pm 0.5,~0.1\pm 0.006$) for (a);($8.05\pm 0.4,0.1\pm 0.005$) for (b); ($8.96 \pm 0.7~,0.2\pm 0.02$) for (c); and ($6.84 \pm 0.8~,0.6\pm 0.07$) for (d). Gray bar length is $\SI{2}{\centi\meter}$.
} \label{fig : general}
\end{figure}

The multiformity in the deformed shapes goes well beyond the occurrence of either a depression or a bump at the bottom surface. The gray level distributions in the mean vertical slices  of (e) or (f) in Fig.~\ref{fig : profils} result from non-axially symmetric shapes.
Views of four three-dimensional reconstructions of representative samples with a set of horizontal slices at various heights (from the top to the bottom of the samples) are presented in Fig.~\ref{fig : general}.
Wrinkles, with a hierarchical structure for samples with the largest $\alpha$, are observed by the top of these samples (Fig.~\ref{fig : general}(a-c)).
The amplitude of the wrinkles as well as the degree of branching is the highest at the top of the shape decreasing downwards until vanishing at the bottom of the hanging cylinder.
In addition, dimples at the lower surface of the samples with the largest aspect ratio are also observed (Fig.~\ref{fig : general}(d)).
Note that no detachment at the upper horizontal plane is observed so that the upper horizontal slice remains circular, with the radius still being that of the mould.

\section{Cascade of wrinkles}

\begin{figure}[!h]
\begin{center}
\includegraphics[width=0.5\textwidth]{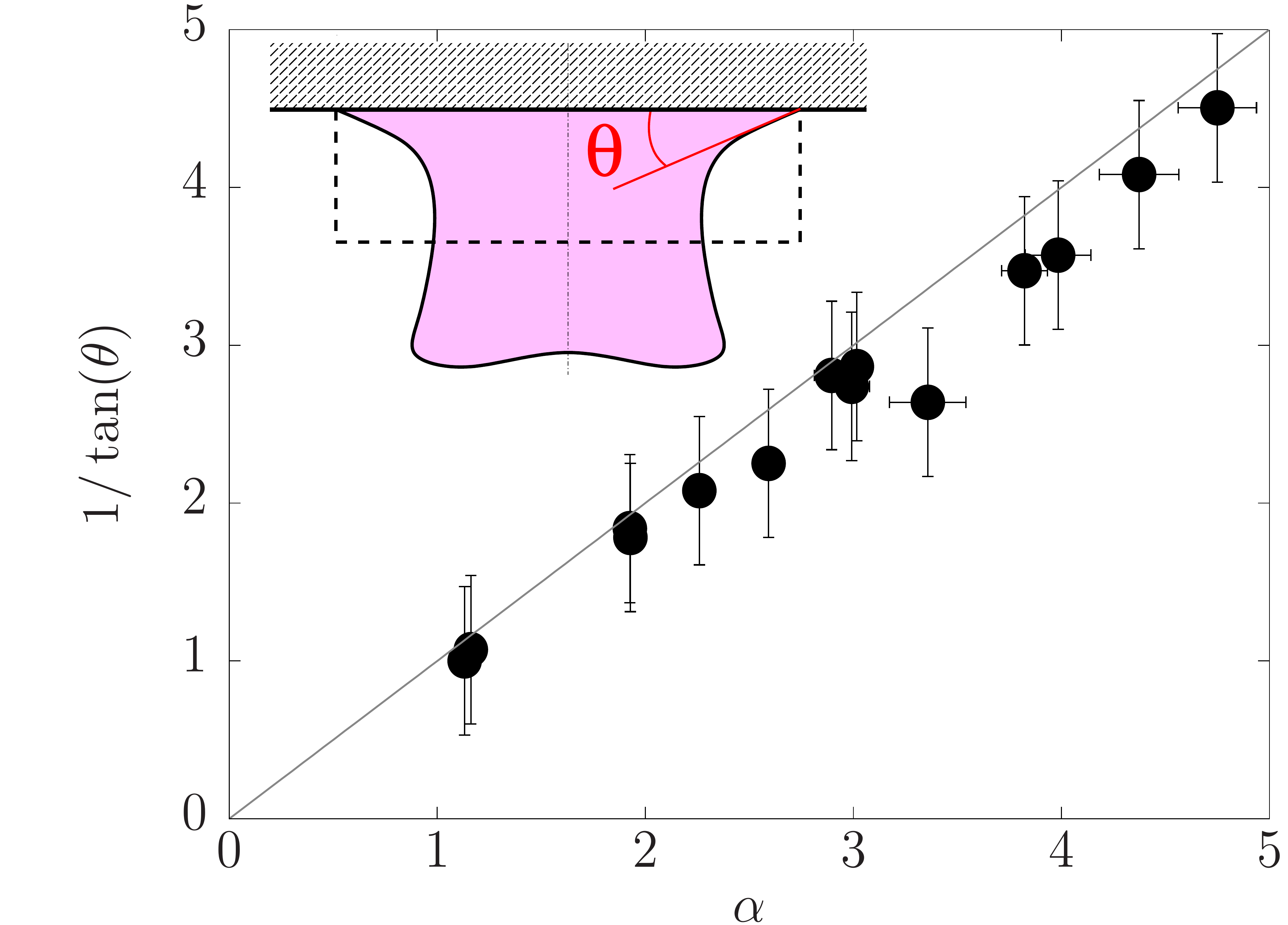}
\end{center}
\caption{$1/\tan \theta$ as a function of the dimensionless acceleration $\alpha$, with $\theta$ the angle formed by the horizontal plate and the tangent of the boundary of the sample in an axial section, measured at the contact line from images obtained with micro-tomography  (for axially symmetric samples). {\bf Inset:} Definition of angle $\theta$.} \label{fig : angle}
\end{figure}

Filled symbols in Fig.~\ref{fig : length} indicate axially symmetric deformed shapes while empty symbols are for non axially symmetric shapes (with wrinkles and/or  dimples).
The separation from axially to non axially symmetric shapes is found to occur for a given value of the dimensionless acceleration, $\alpha \simeq 4.7$. Upon increasing $\alpha$, the second generation in the wrinkles hierarchy of size appears in samples with $\alpha>6.02$.
Let $\theta$ be the angle formed by the tangent of the profile in an axial section and the horizontal direction in this section at the top of the sample (see the scheme in Fig.~\ref{fig : angle}). $\theta$ has been measured for axially symmetric hanging cylinders only (this angle is not defined for samples with wrinkles).  From Fig.~\ref{fig : angle}, $1/\tan \theta$ is found to be similar to $\alpha$.
Since the threshold for the formation of the wrinkles is for $\alpha=\alpha_c=4.7$, one concludes that the wrinkles begin to develop for deformations so that $1/\tan \theta>4.7$, \ie $\theta<\theta_c\simeq \SI{0.21}{\radian}$.
Fringe \cite{Lin2016,Lin2017} or fingering \cite{Crosby2000,Ghatak2000,Saintyves2013,Biggins2013} instabilities have been reported in stretched elastic cylinders or layers attached by their two ends to parallel rigid plates.
The analytic model developed in \cite{Biggins2013,Biggins2018} as well as the simulations of \cite{Lin2017} predict that the interface is linearly unstable for stretch ratio such that the angle $\theta$ at the contact line is smaller than $\SI{0.195}{\radian}$. In addition, the bifurcation has been found to be sub-critical \cite{Biggins2013}, hence an effective threshold angle larger than $\SI{0.195}{\radian}$. 
Moreover, supplementary experiments in which our cylinders made in polyacrylamide gels are attached by their ends to two parallel rigid plates and stretched in an extensometer lead to wrinkles, with the cascading structure, that are definitely similar to those observed in the hanging specimens (see Sec.~\ref{sec : stretched}).
It is therefore tempting to associate the formation of the wrinkles with the mechanism of the fringe/fingering instability. This  mechanism, which can be seen as an extension of the original idea of M.A. Biot \cite{Biot1963}, has first been introduced in \cite{Biggins2013}, and can be sketched in the case of hanging cylinders as follows.\\

\begin{figure}[!h]
\begin{center}
\includegraphics[width=0.25\textwidth]{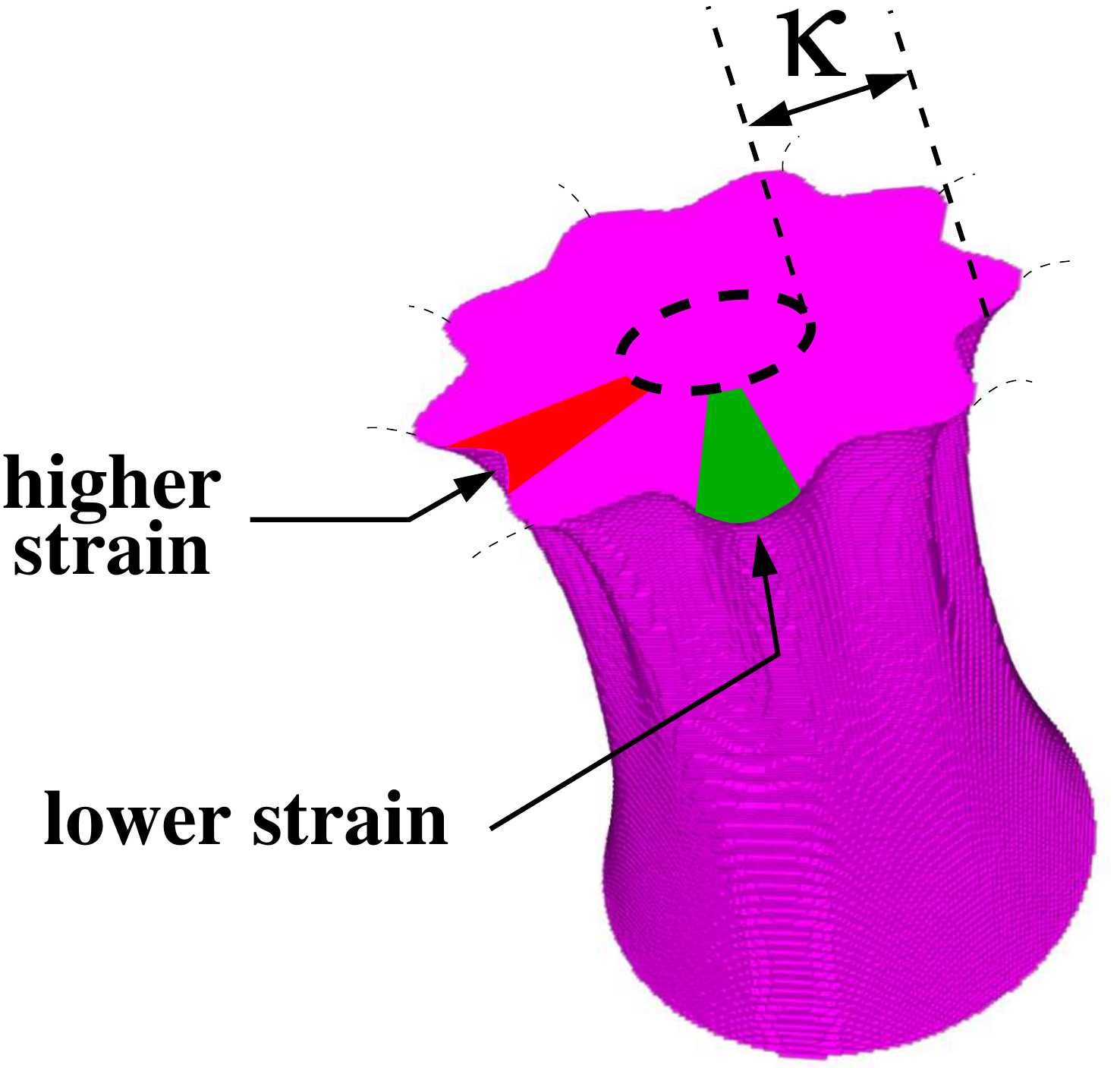}
\end{center}
\caption{Sketch of a piece of a hanging cylinder perturbed by infinitesimal wrinkles. $\kappa$ is the penetration depth (in the radial direction) of the perturbation. The green area in the horizontal section indicates the deformed region near a ridge, while the red one indicates the deformed region near a furrow.}  
 \label{fig : biot}
\end{figure}

Let us start from a hanging elastic cylinder stretched by the gravity and suppose first that the deformed shape is axially symmetric. Consider now a circumferential periodic disturbance consisting of vertical wrinkles below the rigid plate. The characteristic penetration depth of the disturbance of infinitesimal amplitude in the radial direction is finite, laying in between the two relevant geometric length scales which are the wavelength of the disturbance and the radius of the cylinder. The disturbance causes changes in the stretching along the vertical direction: the material is less stretched near the ridges than that it was before the perturbations, whereas it is more stretched near furrows (Fig.~\ref{fig : biot}). The deformed volume involved in a ridge being larger than the volume involved in a furrow, one deduces that the perturbation globally lowers the stretching energy. On the contrary, the additional shear strain induced by the perturbation causes an increase in the strain energy. The competition between the two elastic effects, \ie the reduction in the stretching and the increase in the shearing, is responsible for the instability threshold.\\

\begin{figure}[!h]
\begin{center}
\includegraphics[width=0.35\textwidth]{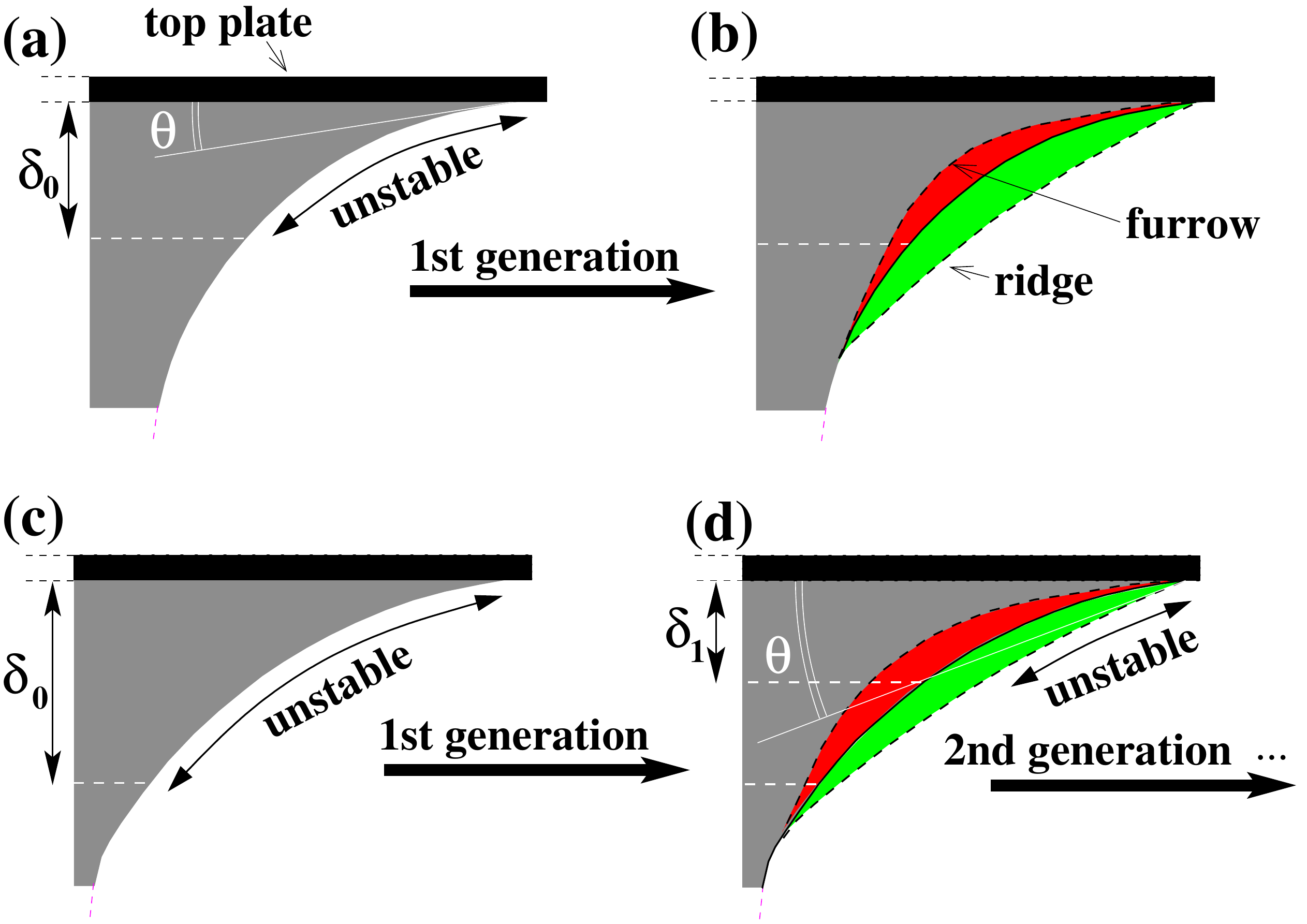}
\end{center}
\caption{Partial sketches of vertical slices, before (a,c) and after (b,d) an infinitesimal perturbation is applied. In (a) the angle $\theta$ (formed by the tangent of the profile at the contact line and the horizontal) is supposed to be smaller than the threshold $\theta_c$. The surface is therefore unstable inside a domain whose depth is $\delta_0$, leading to the formation of ridges and furrows (as sketched in (b)) with a wavelength $\sim \delta_0$.  In (c)  $\alpha$ is larger than in (a,b) so that the angle $\theta$ defined for the ridges is now smaller than $\theta_c$  (d). The top of the ridge is therefore unstable inside a domain of depth $\delta_1$, leading to a new generation of wrinkles (not drawn)  whose wavelength scales as $\delta_1~~(<\delta_0)$}.
 \label{fig : cascading}
\end{figure}

From our experiments, the critical angle $\theta_c$ does not significantly change with the initial dimensions of the samples. This observation, which is consistent with the numerical simulations of \cite{Lin2016}, is an indication that neither the initial radius of curvature nor the initial height play a crucial role at the instability onset, at least for the range of aspect ratios investigated here. Neglecting surface tension, the remaining characteristic length for the instability is the depth $\delta_0$, that demarcates the stable to the unstable parts of the unruffled surface of a given hanging cylinder (see Fig.~\ref{fig : cascading}(a)). The wavelength of the wrinkles is therefore expected to scale as  $\delta_0$, the only relevant characteristic length in the instability mechanism.
Inspired by the mechanism presented in \cite{Pomeau1997,Audoly2010} that accounts for self similar hierarchical wrinkling in constrained thin sheets, we propose now an explanation for the cascading structure observed in Figs.~\ref{fig : general}(a-c).
If $\alpha$ is so strong that the angle $\theta$, defined now at the top of the ridges, would be again larger than $\theta_c$, then the previous ridges are themselves unstable. Let $\delta_1$  be the depth separating the unstable to the stable regions of the ridges (see Fig.~\ref{fig : cascading}(d)). $\delta_0>\delta_1$ because the development of the first generation of wrinkles induces a partial release of the vertical stretch at a ridge.
Since the characteristic wavelength of the resulting secondary wrinkles scales as $\delta_1$, the wavelength of the secondary generation is smaller than the wavelength of the first one. Repeating this argument, we deduce that the wavelength of any new generation scales as the shorter and shorter vertical extension of its wrinkles, in accordance with the hierarchical structure of the wrinkles.

Note that wrinkles leading to transient hierarchical structures reminiscent to those observed here with the polyacrylamide gels, have been observed at the end-plates of filament stretching rheometers during the rapid stretching of viscoelastic fluid filaments \cite{McKinley1996,Hassager2002}. Although the above mechanism, that is based on the minimization of the energy, cannot directly apply to viscoelastic fluids, we think that it can give some insights to better understand this destabilization of stretched viscoelastic filaments.

\section{Depressions as a result of a  Rayleigh-Taylor mechanism}

\begin{figure}[!h]
\begin{center}
\includegraphics[width=0.5\textwidth]{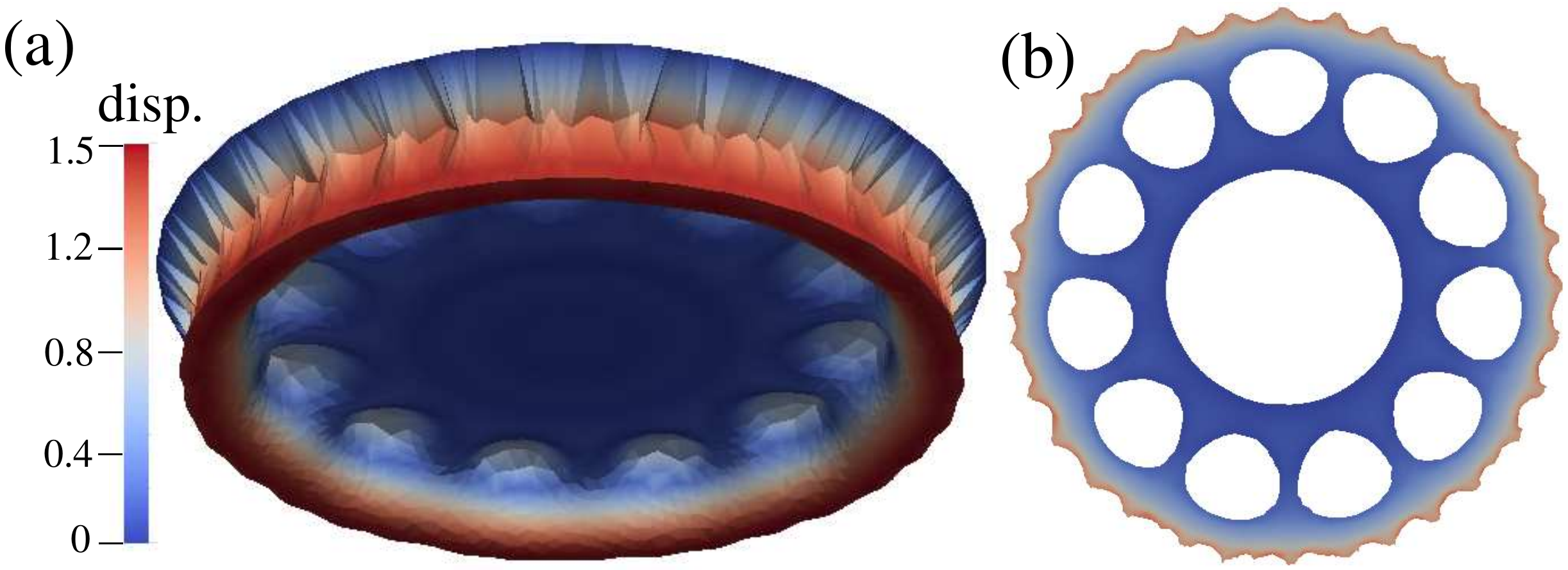}
\end{center}
\caption{ {\bf (a)} Deformed shape computed with $R/h=8$, $\alpha=5.58$ and $\beta=0$. Colors indicate the ratio of the total displacement to the initial height. {\bf (b)} Horizontal slice of the previous deformed shape (a) at the distance $h$ below the rigid plate.} \label{fig : numerics 3d}
\end{figure}

In addition to wrinkles, a massive divergent displacement of the matter from the axis to the periphery of the hanging cylinders leading to the formation of a unique deep central depression is observed in samples with large aspect ratios (see Fig.~\ref{fig : general}(c,d) and Fig.~\ref{fig : profils}(c)). Secondary depressions (dimples) appear for even larger aspect ratios (see Fig.~\ref{fig : general}(d)).
The limit of an ``infinite'' aspect ratio can be approached by considering horizontal elastic infinite layers of height $h$ attached at the top to a rigid substrate, the lower surface being free to deformed. This limit has been studied in the past \cite{Mora_prl2014,Liang2015,Ciarletta2017,Chakrabarti2018}.  When subjected to the gravitational acceleration $g$, the Elastic Rayleigh-Taylor instability makes the downwards facing surface unstable beyond a threshold for the dimensionless acceleration, $\alpha\simeq 5.7$, leading to the formation of deep depressions organized in a hexagonal lattice \cite{Mora_prl2014,Chakrabarti2018}. The underlying mechanism is the competition between gravity (which tends to deform the surface in order to lower the center of mass of the system) and elasticity.
This order of magnitude of this threshold is in agreement with the dimensionless acceleration for which the dimples are observed ($\alpha=6.84\pm 0.8>5.7$).
We have not carried out a systematic study by gradually varying $\alpha$ in order to find its critical value for a given aspect ratio: for shear modulus as low as ten Pascals, any small perturbation in the curing process as a slight unpredictable excess of dioxygen in the reactant, would lead to significant relative changes in the value of the shear modulus. Hence carrying out a series of experiments with targeted low shear modulus is a truly difficult task that would require the production of a considerable amount of trial samples.   
Furthermore, dimensionless surface tension would change with the elastic modulus, and it is no longer negligible for these low shear moduli and small initial heights $h$.
In order to support the scenario of the formation of the dimples as a result of the Elastic Rayleigh-Taylor instability, we have performed 3D Finite element (FE) simulations which consist of minimizing the sum of gravity and neo-Hookean elastic energies of a hanging cylinder (see Sec.~\ref{sec : method}). Surface tension is neglected in these simulations, hence quantitative difference are expected with some of the experiments. The important point is the similarity between the equilibrium shape of sample (d) in Fig.~\ref{fig : general} and the shapes found in the simulations (see Fig.~\ref{fig : numerics 3d}).

\begin{figure}[!h]
\begin{center}
\includegraphics[width=0.5\textwidth]{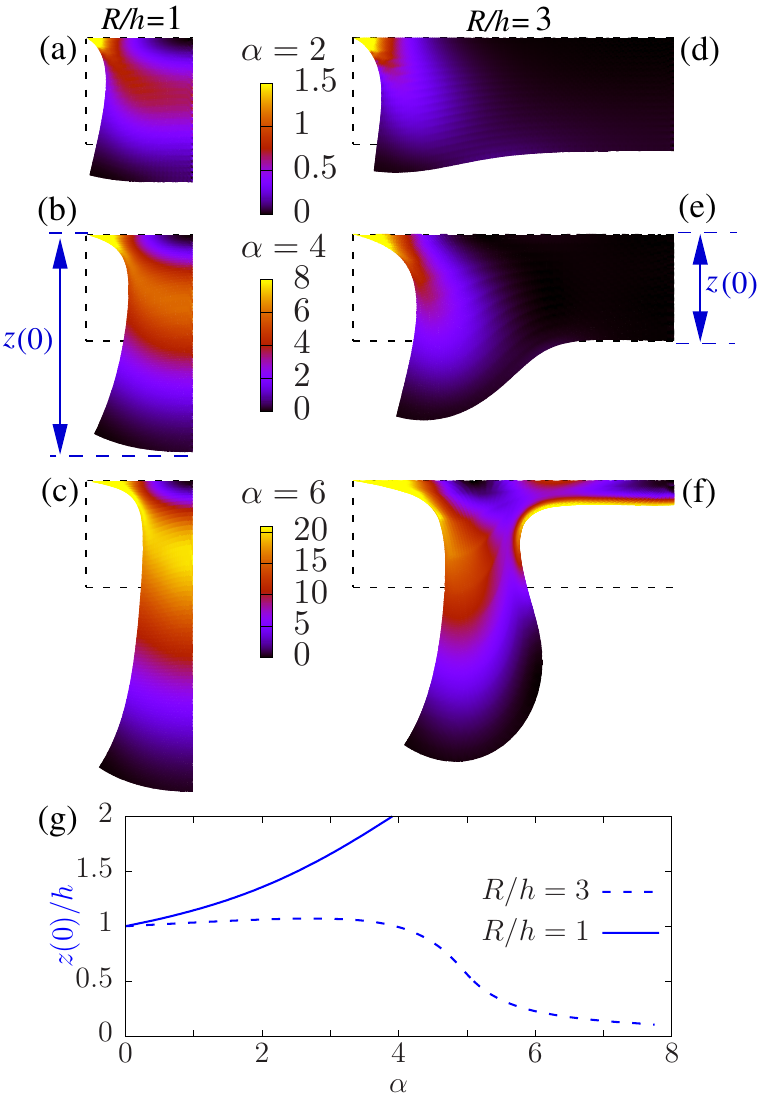}
\end{center}
\caption{Axial sections of hanging cylinders with the aspect ratio $R/h=1$ (a-c) and $R/h=3$  (d-f) for $\alpha= 2$ (a,d), $\alpha=4$ (b,e) and $\alpha=6$ (c,f), computed with the assumption of axial symmetry. Dashed lines indicates the initial shapes.  Color maps indicate the magnitude of the strain energy density divided by $\mu/2$ (formally $\mathrm{tr}(\mathbf{F}\mathbf{F}^T-3)$ where $\mathbf{F}$ is the deformation gradient tensor, see Eq. \ref{eqn : functional}).  (g) Ratio of the thickness of the deformed shape along the axis of the cylinder ($z(0)$, see (b) or (e)) to the initial height ($h$) as a function of $\alpha$, for to aspect ration $R/h$. Dimensionless surface energy is $\beta=0$.} \label{fig : numerics 2d}
\end{figure}

The massive divergent displacement of matter (as in Fig.~\ref{fig : general} (c) and also in Fig.~\ref{fig : profils}(f)) can be viewed as a consequence of the Elastic Rayleigh-Taylor mechanism also.
Supplementary FE simulations have been carried out, again with $\beta=0$ in order to emphasize the competition between gravity and elasticity.
In order to avoid wrinkles that would make the simulations more difficult to handle, the simulations have been carried out by imposing axially symmetric deformations (see Sec.~\ref{sec : 2d}).
The displacements at the top end of the cylinder are fixed to zero, and the other boundaries are free to deform.
Upon gradual increases of the dimensionless acceleration $\alpha$, the remaining thickness below the center of the top cross section is found to first increase, and then for aspect ratios larger than $R/h \simeq 2$, this thickness abruptly decreases beyond in a narrow range of $\alpha$ to reach a value far smaller than the initial height (See Fig.~\ref{fig : numerics 2d}(d-g)).
The evolution of the equilibrium shape can therefore be divided into two regimes: a first regime where the shape evolves progressively as $\alpha$ is increased.
Then beyond a critical $\alpha$, configurations in which the matter is cooperatively displaced downwards by the periphery, start to become more favorable and the rate of deformation as function of $\alpha$ abruptly increases, a mechanism reminiscent to that of the Rayleigh-Taylor instability.
The value of the critical $\alpha$ (\ie $\alpha\sim 4.5$ for $R/h=3$ as in Fig.~ \ref{fig : numerics 2d}(g)) is smaller than the threshold in the Rayleigh-Taylor instability because of the absence of lateral constraints.

\section{Conclusion}

  Since applications with soft materials often rely on their ability to withstand large deformations, hence low elastic modulus, the characteristic dimensionless gravitational acceleration $\alpha$ can be significant. The competition between elasticity and gravity must therefore be considered, with possible dramatic effects, such as those highlighted in this paper.\\ 
Although the system consisting of a hanging elastic cylinder is probably one of the simplest in order to study the mechanical response of highly stretchable elastic material, the resulting deformations are manifold, whose underlying mechanisms have been unveiled in this paper.
Hierarchical wrinkles result from an elastic instability in which the role of the gravity is just to fix the deformation of the unperturbed state (since they also develop in the stretched geometry in which gravity is not a key parameter). {The cascading structure arising from this instability has been evidenced, and explained schematically from basic physical arguments.
In contrast, the large central depression as well as the dimples result from the competitive interplay between elasticity and gravity. This competition was known to drive the formation of hexagonal patterns as a result of the elastic Rayleigh-Taylor instability. By considering the hanging cylinder geometry, we have demonstrated another effect arising from this competition, consisting of a dramatic changes in the global shape of the deformed samples, ranging from convex (convergent displacement of the matter towards the axis) to highly concave (formation of a large and deep central depression).     
A detailed analytic description of these phenomena would first require analytic expressions for the base deformed shapes prior to the occurrence of these patterns.  This is a difficult task due to non-linearities induced by the large deformations in the system.\\

This work is focused on the equilibrium configurations of soft gels, for which the Helmholtz energy is maximal. Further works dealing with dynamical processes, as those involved in the transition from one equilibrium shape to the another one as a result of a abrupt change of the volume force,  or in the cutting of such soft materials when they are strongly strained, would be an interesting continuation of this study and would require to take into account dissipative processes in addition to elasticity.\\

X-ray micro-tomography has proved to be a powerful technique capable of capturing complex equilibrium shapes of hydrogels subjected to large deformations, whereas conventional optical methods are more difficult to handle with these transparent and folded samples.\\

SM thanks S.Lin for enlightening discussions about simulations and computations published in \cite{Lin2017}.   


\appendix
\renewcommand\thefigure{\thesection.\arabic{figure}}
\section{Numerical simulations of hanging cylinders}
This appendix deals with the implementation of simulations of hanging cylinders and to comparisons with the experiments. The numerical method based on the finite element (FE) method, is introduced in Sec. \ref{sec : method}. It consists of the minimization of the augmented energy of a system composed of an incompressible neo-Hookean cylinder bounded to an upper plane and subjected to the combined action of gravity and surface tension. The simulations are used to compute the deformed shapes of hanging cylinders presented in Fig.~6(d,e) of the main article, with the goal to show that the interplay between elasticity and gravity can explain the formation of secondary depressions (dimples) as those observed in the experiments.
The deformed shape of hanging cylinders with the same parameters (geometry, mass density, gravity, shear modulus and surface tension) as in the experiments reported in the main article are also computed in Sec. \ref{sec : 2d}. These simulations are made by fixing the azimuthal displacements to zero, an approximation which amounts to considering a mean displacement as in the main average slices obtained by micro-tomography.   

\subsection{Numerical method} \label{sec : method}

We use numerical computations based on the FE method to predict the equilibrium configuration of an elastic cylinder (of radius $R$ and height $h$) under the action of the gravity and capillary forces, with the condition of zero displacement at the upper end.
We model the body as an incompressible neo-Hookean solid (of shear modulus $\mu$) with a surface energy proportional to the total area of its boundary in the current configuration  (coefficient of proportionality $\gamma$). We accept that the body can undergo large displacements and deformations, by using a fully non-linear kinematical theory.

Let $\Omega$ be the cylindrical domain (height $h$ and radius $R$) in the reference configuration, $\mathbf{r}$ be the material point location in this reference configuration,  $\mathbf{u}(\mathbf{r})$ be the displacement vector, and $\mathbf{F}(\mathbf{r})$ be the deformation gradient tensor (formally, $\mathbf {F}=\partial (\mathbf{r}+\mathbf{u})/\partial\mathbf{r}$).
To characterize equilibrium configurations, we consider the augmented energy
\begin{equation}
  {\cal E}(\mathbf{u},p)=
  \int_\Omega\left(
  \frac{\mu}{2}\left(\mathrm{tr} (\mathbf{F}\mathbf{F}^T)-3\right)
  -\rho \mathbf{g} \cdot  \mathbf{u}
  +\mu p(J-1)
  \right)\mathrm{d}\mathbf{r}
  +\gamma \int_{\partial\Omega} \left\Vert J\,\mathbf{F}^{-T}\mathbf{N}\right\Vert\,\mathrm{d}S.
\label{eqn : functional}
\end{equation}
The first term in the first integral is the strain energy density for an isotropic and incompressible neo-Hookean solid. The second term is the gravity energy density. The Lagrange multiplier $p(\mathbf{r})$ ensures the incompressibility constraint $J= 1$.
The second integral accounts for the surface energy, with vector $\mathbf{g}$ the acceleration due to gravity at the Earth's surface. The term $\left\Vert J\,\mathbf{F}^{-T}\mathbf{N}\right\Vert\,\mathrm{d}S$ gives the element of area of the deformed configuration as a function of the quantities defined on the reference domain $\Omega$, $\mathbf{N}$ being the unit normal to the reference boundary $\partial \Omega$ \cite{Ogden1984,Mora_prl2013}.\\

Eq.~\ref{eqn : functional} can be expressed in terms of the dimensionless quantities $\alpha=\rho g h/\mu$ and $\beta=\gamma/(\mu h)$ by considering the transformation $\mathbf{r}=h\mathbf{r'}$, $\mathbf{u}=h\mathbf{u'}$: 
\begin{equation}
  \frac{{\cal E}(\mathbf{u'},p)}{\mu h^3}=
  \int_{\Omega'}\left(
  \frac{1}{2}\left(\mathrm{tr} (FF^T)-3\right)
  -\alpha \mathbf{u'\cdot e_z}
  +p(J-1)
  \right)\mathrm{d}\mathbf{r'}
  +\beta \int_{\partial\Omega'} \left\Vert J\,\mathbf{F}^{-T}\mathbf{N}\right\Vert\,\mathrm{d}S',
\label{eqn : functional dimensionless}
\end{equation}
with $\Omega'$ a cylinder of height $1$ and radius $R/h$. $\mathbf{e_z}$ is the downwards vertical unit vector.  

Our FE formulation is based on the research of the stationary points of the total energy functional given by Eq.~\ref{eqn : functional dimensionless}. An ad-hoc numerical code is developed for the purpose using the \texttt{FEniCS} finite element library \cite{Fenics2012}. The displacement vector $\mathbf{u'}$ and the Lagrange multiplier $p$ are discretized using Lagrange FEs with a quadratic interpolation for $u'$ and a linear interpolation for $p$, on a triangular mesh (see ch. 20 of [26]). The linear mesh density in either direction is written $n_{mesh}$.
The nonlinear problem in the ($u'$, $p$) variables is solved using a Newton algorithm based on a direct parallel solver (\texttt{MUMPS}). 

Quasi-static simulations are performed by progressively increasing the surface tension $\beta$ up to the desired dimensionless value, then by progressively incrementing the load parameter $\alpha$, recording the displacement field and the Lagrange multiplier, and reaching convergence at each step.

\subsection{FE simulations with imposed axially symmetric deformation} \label{sec : 2d}

A fine mesh density would be required in order to account for the formation of wrinkles, hence prohibitive computational times. Here, our strategy is to ignore in the simulations any non axially symmetric deformation by considering them to be a perturbation of an axially symmetric base deformation.
The shape of this  base (axially symmetric) deformation is computed and compared with the mean vertical slices obtained by micro-tomography. 

The displacement vector is expressed in a cylindrical coordinate system as $\mathbf{u}=u_r(r,z)\mathbf{e_r}+u_z(r,z)\mathbf{e_z}$, hence a two-dimensional problem.  The deformation gradient tensor to be inserted in Eq.~\ref{eqn : functional dimensionless} is then
\begin{equation}
  \mathbf{F}=\left( \begin{array}{ccc}
    1+\partial u_r/\partial r,0,\partial u_r/ \partial z\\
    0,1+u_r/r,0\\
    \partial u_z/\partial r,0,1+\partial u_z / \partial z
    \end{array} \right).
\end{equation}

A FE simulation (with enforced axial symmetry) has been carried out for each samples whose shape has been acquired by micro-tomography.
The shapes, calculated from FE simulations, are superimposed with the mean vertical slices determined by micro-tomography in Fig.~\ref{fig : axially symmetric}. In order to facilitate the visualization of the junction between the gel and the rigid  upper plate, a homogeneous mask was added after images reconstruction at the precise location of the upper plate.  Because elastic bodies with low elastic modulus are sensitive to any slight disturbance, the deformed samples are never perfectly axially symmetric and one may observe thin areas close to the boundary of the deformed cylinder with a progressive change in the gray level, from light to dark gray. In addition, deformed samples which are non axially symmetric due to wrinkles or dimples lead to a continuous distribution of the gray level at points with coordinates ($r,z$) that can be either inside or outside the sample depending on the azimuth.

In the simulations, the dimensionless acceleration $\alpha$  has been directly calculated from the value of $\mu$ determined by indentation tests, and the dimensionless surface tension $\beta$ is calculated by taking for $\gamma$ the surface energy coefficient of water since the gels are mainly composed of water (more than 99.9\% in mass). 

\begin{figure}[!t]
\begin{center}
  \includegraphics[width=\textwidth]{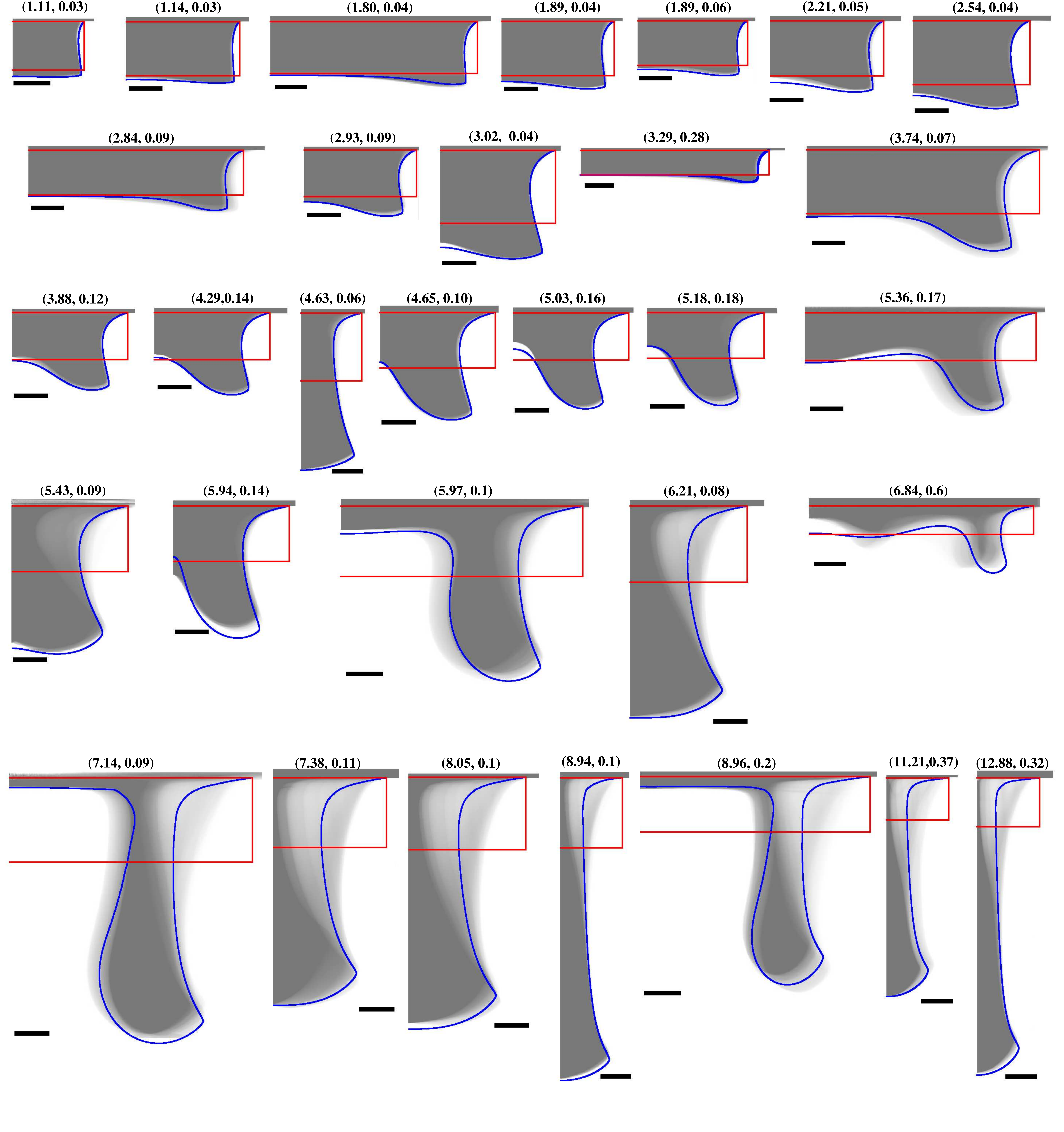}
  \caption{Vertical slices of hanging cylinders captured by micro-tomography, obtained by averaging over the azimuth. The left border of each image is the axis of revolution of the initial (undeformed) shape, whose initial dimensions are indicated by the red lines.  Blue lines reproduce the deformed shapes computed by FE simulations with $n_{\mathrm{mesh}}=20$, for which $\alpha$ and $\beta$  are indicated at the top of each image. Bar length is 1 cm.} 
  \label{fig : axially symmetric}
\end{center}
\end{figure}

The comparison between experiments and numerics is satisfactory (Fig.~\ref{fig : axially symmetric}). The observed discrepancies result from the uncertainties in the determination of $\mu$, and from slight defects in the samples. The physical ingredients used in the simulations (the neo-Hookean constitutive law, with gravity and surface tension) are therefore enough to quantitatively capture the averaged shape of the hanging cylinders. In cases of samples with wrinkles or secondary depressions (dimples), the good agreement between the mean vertical slices and the axially symmetric FE simulations, support the view of wrinkles or dimples as perturbations of a axially symmetric base states.\\

In this appendix, a numerical method has been introduced. It has been used to show that the interplay between gravity and elasticity produces dimples at the bottom interface of hanging cylinders with a large aspect ratio. In addition, enforcing axially symmetric deformations in the simulations yields to deformed shapes that well reproduce the mean vertical slices obtained in the experiments.

\section{Stretched cylinder} \label{sec : stretched}

This appendix deals with an experiment carried out with a cylinder made in polyacrylamide gel, attached by its ends to two parallel rigid plates and stretched in an extensometer.

The initial radius is $r= \SI{7.2}{\centi\meter}$ and the initial height is  $ h=\SI{2}{\centi\meter}$. The shear modulus of the polyacrylamide hydrogel is $\mu=\SI{51.8 \pm 1.5}{\pascal}$. The elasto-capillary length, defined as the ratio of the surface tension ($\gamma \sim 70$ mN/m) to the shear modulus, is $\ell_{ec}\sim 1.3$ mm. 

The sample is stretched step by step by the extensometer with increments of $ \SI{3}{\milli\meter}$. For each step, the equilibrium shape is acquired using an x-ray scanner. The resolution is adapted so that the size of the cubic voxels is $(\SI{200}{\micro\meter})^3$.

Representative three-dimensional reconstructions of the deformed cylinder are show in Fig.~\ref{fig : stretch} for three different stretch ratio.  Fig.~\ref{fig : stretch}-(a) shows the sample in an almost unstretched configuration. The lateral surface is not straight because of the effect of gravity.  In Fig.~\ref{fig : stretch}-(b) the stretch ratio is lower than the threshold reported in \cite{Biggins2013,Lin2017} for the fingering instability and the sample is, accordingly, axially symmetric. 
In Fig.~\ref{fig : stretch}-(c) the threshold has been reached, and one recovers the fingering instability reported in \cite{Lin2016,Lin2017}. In Fig \ref{fig : stretch}-(d) the stretch ratio is larger and one observes the formation of new generations of wrinkles.  

The outer boundaries of the horizontal slices on the right of  Fig.~\ref{fig : stretch} share common features with horizontal slices in Fig.~3 of the main article. Note that the characteristic lengthscales in the observed patterns are larger than the elasto-capillary length $\ell_{ec}$, hence capillary effects are not expected to be significant.

\begin{figure}[!h]
\begin{center}
\includegraphics[width=0.45\textwidth]{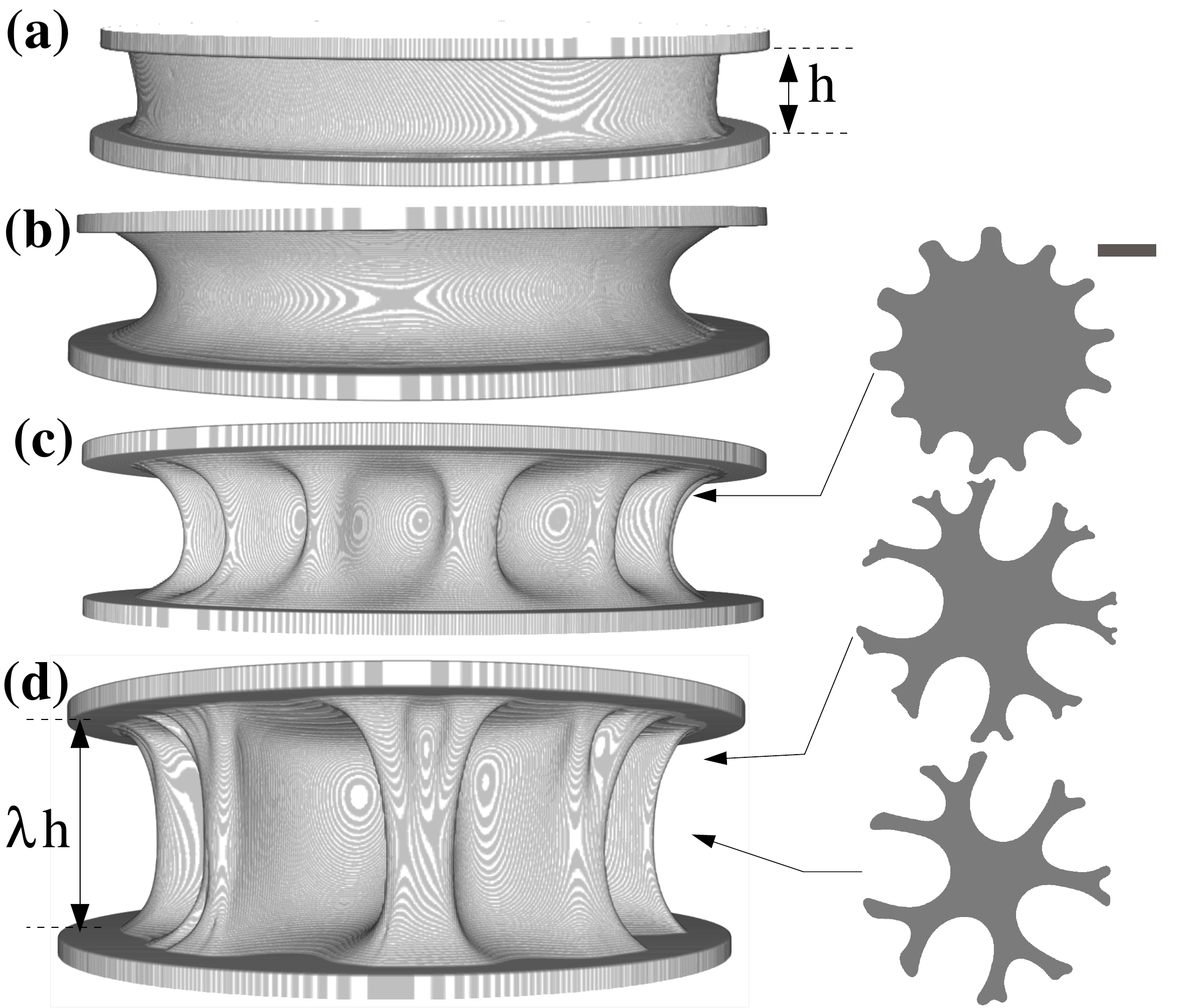}
\end{center}
\caption{Three dimensional views and horizontal slices obtained with micro-tomography of a cylindrical layer of hydrogel (shear modulus $\mu=\SI{51.8\pm 1.5}{\pascal}$; initial dimensions $R\times h=\SI{7.2}{\centi\meter} \times \SI{2}{\centi\meter}$) ; {\bf (a)} in an almost unstretched configuration with the stretch ratio $\lambda=1.05$  ;  {\bf (b)} with the stretch ratio $\lambda=1.3$ ; {\bf (c)} with the stretch ratio $\lambda=1.6$ ; {\bf (d)} with the stretch ratio $\lambda=2.5$. The position of the horizontal slices in the gels are indicated by the arrows. Bar length is equal to $\SI{2}{\centi\meter}$ (for the horizontal slices).} 
 \label{fig : stretch}
\end{figure}

\end{document}